\input amstex 
\documentstyle{amsppt}\NoBlackBoxes
\pagewidth{30pc} \pageheight{47pc} 
\magnification = 1200
 
\input epsf.tex

\def\today{\number\day\space\ifcase\month\or January\or February\or
March\or April\or May\or June\or July\or August\or September\or
October\or November\or December\fi\space\number\year}
 
\newcount\tagc \global\tagc=0
\def\tagno{\global\advance\tagc by 1 \tag\number\tagc}
\newcount\headc \headc=0
\def\headno{\global\advance\headc by 1 \number\headc . \global\propc=0}
\newcount\propc \propc=0
\def\propno{\global\advance\propc by 1 \number\headc .\number\propc}
\newcount\figc \figc=1
\def\figno{\botcaption{Figure \number\figc}
\endcaption \global\advance\figc by 1}
 
\def\nextfigure#1{\edef#1{\number\figc}}
\def\name#1{\xdef#1{{\number\headc .\number\propc}}}

 \def\End{\operatorname{End}}
\def\Hom{\operatorname{Hom}}

\def\tr{\operatorname{tr}}

\def\C{{\Bbb C}}

\def\Z{{\Bbb Z}}

\define\ass{{\alpha}}
\define\pdual{{\gamma}}
\define\nat{{\tau}}
\define\pv{{\nu}}
\define\ring{{\Bbb F}}

\topmatter
\title Spherical categories \endtitle
\author John W. Barrett \& Bruce W. Westbury \endauthor
\date 10 August 1993; revised 22 July 1998\enddate
\address
School of Mathematical Sciences,
University of Nottingham,
University Park,
Nottingham,
NG7 2RD, UK
\endaddress
\email jwb\@maths.nott.ac.uk \endemail
\address
Mathematics Institute,
University of Warwick,
Coventry,
CV4 7AL, UK
\endaddress
\email bww\@maths.warwick.ac.uk \endemail
\toc
\head {0.} Introduction \endhead
\head {1.} Monoidal categories with duals\endhead
\head {2.} Spherical categories \endhead
\head {3.} Spherical Hopf algebras \endhead
\endtoc

\abstract
This paper is a study of monoidal categories with duals where the tensor
product need not be commutative. The motivating examples are categories
of representations of Hopf algebras. We introduce the new notion of a
spherical category. In the first section we prove a coherence theorem 
for a monoidal category with duals following \cite{MacLane  1963}.
In the second section we give the definition of a spherical category,
and construct a natural quotient which is also spherical.
In the third section we define spherical Hopf algebras so that the category
of representations is spherical. Examples of spherical Hopf algebras are 
involutory Hopf algebras and ribbon Hopf algebras. Finally we study the
natural quotient in these cases and show it is semisimple.
\endabstract
\endtopmatter

\document
\head {0.} Introduction \endhead

In this paper we introduce the notion of a spherical category. The
motivation for this and the main application is to generalise the
Turaev-Viro state sum model invariant of a closed piecewise-linear
3-manifold. This application is discussed separately in \cite{Barrett and Westbury 1996}.

A spherical category is defined to be a pivotal category which satisfies
an extra condition, which states that two trace maps are equal. 
The universal strict pivotal category has an interpretation
in terms of planar graphs. A number of authors have introduced
categories whose morphisms are defined in terms of diagrams and which
have applications to low dimensional topology (for example,
\cite{Freyd and Yetter  1989}, \cite{Joyal and Street  1991}, \cite{Reshetikhin and Turaev  1990}).
The main difference between pivotal and spherical categories and these
categories is that pivotal and spherical categories are not required to be
braided. However, any braided pivotal category satisfies the extra spherical
condition. The geometrical interpretation of the extra condition, and the
reason for the terminology, is that in a pivotal category planar graphs
are equivalent under isotopies of the plane whereas in a spherical category
closed planar graphs are equivalent under isotopies of the 2-sphere.

The first part of the paper gives the definition of a monoidal category 
with duals and the proof of a coherence theorem. The proof follows the proof
of the coherence theorem for monoidal categories in \cite{MacLane  1963}. 
A monoidal category with duals can be thought of
as an abstraction of the category of representations of a Hopf algebra,
except that we require each object to be reflexive.
A coherence theorem for categories with duals is given in \cite{Kelly and Laplaza  1980}.
The differences between this result and ours are that we do not assume that
the monoidal category is symmetric but we do assume that all objects are
reflexive. The difference between their proof and ours is that they emphasise
the adjunctions involving $\Hom$ and $\otimes$ whereas we emphasise the
contravariant duality functor (following \cite{Freyd and Yetter  1989}). The reason we
need this result is that the category of representations of a Hopf algebra
is not strict.

The main sources of spherical categories are categories of representations
of Hopf algebras with a distinguished element satisfying conditions. These
are called spherical Hopf algebras. The two main examples of such Hopf
algebras are involutory Hopf algebras and ribbon Hopf algebras (such as
quantised enveloping algebras). The reason we consider the more general
notion of a spherical category instead of spherical Hopf algebras is that
in the applications to 3-manifold invariants it is necessary to pass to a
quotient. This quotient is not the category of representations of a Hopf
algebra. The definition of this quotient applies to any additive
spherical category. 

In the last section we show that the quotient of the category of 
representations of a spherical Hopf algebra is semisimple. This is the
main result in the paper. The proof uses properties of the matrix trace
and does not apply to the quotient of an arbitrary additive spherical
category. Although this quotient is semisimple the set of isomorphism
classes of simple objects may be infinite even if the original Hopf
algebra is finite dimensional. The construction of 3-manifold invariants
depends on finding a spherical subcategory such that, in the semisimple
quotient of this subcategory, the set of isomorphism classes of simple objects
is finite. For each quantised enveloping algebra at a root of unity such a
subcategory can be constructed using the results in \cite{Andersen  1992} and \cite{Andersen and Paradowski 1995}.

\head\headno Monoidal categories with duals \endhead

In this section we study monoidal categories with duals. Examples are
finitely generated free modules over a commutative ring. In these examples
the tensor product is commutative, but we do not require this. The
motivating examples of monoidal categories with duals are the categories
of representations of Hopf algebras. These categories have a contravariant
functor given by the contragradient dual but are not categories with duals,
in general, as we require that each object be reflexive.

The only result in this section is a coherence theorem. The proof follows
the coherence theorem for monoidal categories in \cite{MacLane  1963}.
The definition of coherence involves an infinite set of diagrams. A
coherence theorem gives a finite set of conditions for coherence to hold.

In the following definition it is convenient to regard the distinguished
object $e$ as a functor from the category with one morphism to $\Cal C$. 
Then $\lambda$ and $\rho$ can be defined to be natural transformations.

\definition{Definition \propno} A monoidal category is a category, $\Cal C$,
together with a functor $\otimes\colon{\Cal C}\times{\Cal C}\to {\Cal C}$
an object $e$, and natural transformations
$$\align
\ass\colon (\otimes\times 1)\otimes &\to (1\times\otimes )\otimes\\
\lambda\colon e\otimes 1&\to 1\\
\rho\colon 1\otimes e&\to 1\endalign$$

Each component of each of the natural transformations $\ass$, $\lambda$ 
and $\rho$ is required to be an isomorphism. This data is also required
to be coherent which is equivalent to requiring that $\ass$ satisfy the 
pentagon identity and that $\rho$ and $\lambda$ satisfy a triangular 
identity. Also, $\rho(e)=\lambda(e)\colon e\otimes e\to e$.
These conditions are given in \cite{MacLane  1971}.                               
          
If each component of the natural transformations $\ass$, $\lambda$ and 
$\rho$ is the identity map then the category is called strict monoidal.
\enddefinition

If $C$ is a category, then the opposite category $C^-$ is the category 
obtained by reversing morphisms. If $F\colon C\to D$ is a functor, then
$F^-\colon C^-\to D^-$ is the corresponding functor on the opposites. A
functor $C^-\to D$ is the same as a contravariant functor $C\to D$.

\definition{Definition \propno} 
Let $({\Cal C},\otimes, e, \ass, \lambda, \rho)$ be a monoidal category.
Then the data needed to make this a category with duals is a  
functor, $*\colon {\Cal C^-}\to {\Cal C}$, and natural transformations 
\roster
\item $\nat\colon 1\to **$,
\item $\pdual\colon (*\times *)\circ\otimes\to\otimes^{op}\circ *$
\endroster
and an isomorphism $\pv\colon e\to e^*$.

Each component of the natural transformations $\nat$ and $\pdual$ is
required to be an isomorphism. A category with data satisfying these
conditions is called a category with dual data.
\enddefinition

\definition{Definition \propno}  
A category with dual data is strict if each component of the natural
transformations $\pdual$ and $\nat$ is the identity map and the
map $\pv$ is the identity map. In this case the category is called
a category with strict dual data.
\enddefinition

In the following definition, if $\alpha=(\pm,\pm,\ldots,\pm)$ is an $n$-tuple
of signs, then $\Cal C^\alpha$ denotes the category $\Cal C^\pm\times 
\Cal C^\pm\times \ldots\times \Cal C^\pm$, with $\Cal C^+\equiv \Cal C$. 
Also, if 0 is the 0-tuple, then  ${\Cal C}^0$ is
the category with one morphism, so that $e\colon{\Cal C}^0\to{\Cal C}$.

It is convenient to regard $\nu$ as a natural transformation from the 
functor $e$ to the functor $e^-\circ *$.

\definition{Definition \propno} The set of iterates of the functors
$\otimes$, $*$ and $e$ is the smallest set of functors 
$F\colon{\Cal C}^\alpha\to{\Cal C}$ which satisfies the following conditions:
\roster
\item The functor $e\colon{\Cal C}^0\to{\Cal C}$ is an iterate.
\item The identity functor $1\colon{\Cal C}\to{\Cal C}$ is an iterate.
\item If $F$ and $G$ are iterates then the functor $F\otimes G=F\times 
G\circ\otimes$ is an iterate.
\item If $F$ is an iterate then $F^*=F^-\circ *$ is an iterate.
\endroster

The  length of $\alpha$ is called the multiplicity of the iterated functor.
There is a category with objects the set of all
iterates of $\otimes$, $*$ and $e$ and morphisms all natural transformations.
\enddefinition 

\definition{Definition \propno} Let $F$, $G$ and $H$ be any three
iterated functors. Then we have natural transformations
$$\align
\ass\colon (F\otimes G)\otimes H &\to F\otimes (G\otimes H)\\
\lambda\colon e\otimes F &\to F \\
\rho\colon F\otimes e&\to F \\
\pdual\colon (F^*)\otimes (G^*) &\to (G\otimes F)^* \\
\nat\colon F &\to F^{**} \\
\pv\colon  e &\to e^*
\endalign$$
Each of these is an isomorphism. Each of these and their inverses are called
instances. More particularly, the first one, for example, is called an 
instance of $\ass$, and its inverse an instance of the inverse of $\ass$.
\enddefinition 

\definition{Definition \propno} The set of expansions is the smallest
subset of the set of natural transformations which satisfies the following:
\roster
\item The identity functor, 1, is an expansion.
\item Any instance is an expansion.
\item If $\beta$ is an expansion then $\beta\otimes 1$ and $1\otimes\beta$
are expansions.
\item If $\beta$ is an expansion then $\beta^*$ is an expansion.
\endroster
The smallest set which satisfies all but the last condition is the set
of reduced expansions.
\enddefinition

\definition{Definition \propno} The smallest subcategory which contains 
instances and which is closed under $\otimes$ and $*$ is equal to the smallest 
subcategory which contains expansions. This is called the category of iterates 
and any morphism in this category is called an iterated natural transformation.

\noindent The smallest subcategory of the category of iterates which contains 
instances and which is closed under $\otimes$ is equal to the smallest 
subcategory of the category of iterates which contains reduced expansions. Any 
morphism in this category is called a reduced iterated natural
transformation. \enddefinition 

\definition{Definition \propno} Since every expansion is an isomorphism
and the inverse of any expansion is an expansion it follows that
every iterated natural transformation is an isomorphism and that every
inverse of an iterated natural transformation
is an iterated natural transformation. The category ${\Cal C}$ with dual data
is coherent if any two iterated natural transformations between the same
two iterated functors are equal. Equivalently, the dual data is coherent if
any diagram with vertices iterated functors and arrows expansions commutes.
\enddefinition
A category with dual data which is coherent is called a category with duals.

The following is a coherence theorem, in the sense that this is a finite
set of conditions on the data which are necessary and sufficient
for the data to be coherent.

\proclaim{Theorem \propno \name\coherence} A category ${\Cal C}$ with
dual data is coherent if, and only if, the following conditions are 
satisfied:
\roster
\item $\nat (e)$ is the composite
$$e @>{\pv}>> e^* @>\pv^{*-1} >> e^{**}$$
\item For each object $c$, $\nat (c^*)$ and $\nat (c)^*$ are
inverse isomorphisms.
\item For any object $a$, the following diagram commutes
$$\CD
a^*\otimes e @>{1\otimes\pv}>> a^*\otimes e^*\\
@V{\rho (a^*)}VV @VV{\pdual (a,e)}V\\
a^* @>>{\lambda^*(a)}> (e\otimes a)^*
\endCD$$
\item For any object $a$, the following diagram commutes
$$\CD
e\otimes a^* @>{\pv\otimes 1}>> e^*\otimes a^*\\
@V{\lambda (a^*)}VV @VV{\pdual (e,a)}V\\
a^* @>>{\rho^*(a)}> (a\otimes e)^*
\endCD$$
\item For all objects $a$, $b$ and $c$, the following diagram commutes 
$$\CD
(c^*\otimes b^*)\otimes a^*
@>{\pdual (c,b)\otimes 1}>>
(b\otimes c)^*\otimes a^* 
@>{\pdual (b\otimes c,a)}>>
(a\otimes (b\otimes c))^* \\
@V{\ass (c^*,b^*,a^*)}VV
@. 
@VV{\ass^*(a,b,c)}V \\
c^*\otimes (b^*\otimes a^*)
@>{1\otimes\pdual (b,a)}>>
c^*\otimes (a\otimes b)^* 
@>{\pdual (c,a\otimes b)}>>
((a\otimes b)\otimes c)^* 
\endCD$$
\item For all objects $a$ and $b$ the following diagram commutes
$$\CD
a\otimes b @>>{\nat (a\otimes b)}> (a\otimes b)^{**}\\
@V{\nat (a)\otimes \nat (b)}VV @VV{\pdual^*(b,a)}V\\
a^{**}\otimes b^{**} @>{\pdual (a^*,b^*)}>> (b^*\otimes a^*)^*
\endCD$$
\endroster
\endproclaim

Note that the following diagram shows that $\nat (a)^{**}=\nat (a^{**})$
for any object $a$ follows from the condition that $\nat$ is a natural
transformation with each component an isomorphism.
$$\CD
a @>{\nat (a)}>> a^{**}\\
@V{\nat(a)}VV @VV{\nat^{**} (a)}V\\
a^{**} @>>{\nat (a^{**})}> a^{****}
\endCD$$
This is implied by condition (2) but not conversely.

It is trivial that if the natural transformations are coherent then
each of these diagrams commutes. It remains to prove the converse.

\proclaim{Lemma \propno} Any iterated natural transformation is equal to
a reduced iterated natural transformation.
\endproclaim

\demo{Proof} There are six natural transformations $\ass$, $\lambda$,
$\rho$, $\pdual$, $\nat$ and $\pv$ and six conditions above. Furthermore
each condition has exactly one instance which is not reduced and for each 
natural transformation there is exactly one condition which contains an 
instance of that natural transformation which is not reduced. Hence, by 
applications of these diagrams and their images by $*$, any 
instance is equal to a reduced iterated natural transformation and it follows 
that any iterated natural transformation is equal to a reduced iterated 
natural transformation. \enddemo 

This shows that it is sufficient to prove the coherence theorem for
reduced iterated natural transformations. Any reduced iterate of $(\ass, 
\lambda, \rho)$ and their inverses is called a central iterate. Maclane proves 
the coherence theorem for central iterates in \cite{MacLane  1963}. 

\proclaim{Lemma \propno \name\directed} Any expansion of an instance of 
$\pdual$, $\nat$, or $\pv$ (but not their inverses) composed with a central 
iterate is equal to a central iterate composed with a expansion of an instance 
of $\pdual$, $\nat$, or $\pv$. \endproclaim 

\demo{Proof} This follows from the condition
that the central iterate is a natural transformation and that each
component of the expansion is a morphism. The commutative square which 
contains both of these provides the proof.
\enddemo

An example is provided by the following commutative diagram:
$$\CD
((F^*\otimes G^*)\otimes H)\otimes J @>\ass(F^*\otimes G,H,J)>>(F^*\otimes 
G^*)\otimes(H\otimes J)\\
@V(\gamma(F,G)\otimes 1_H)\otimes 1_J VV @VV \gamma(F,G)\otimes 1_{H\otimes J} 
V\\
((G\otimes F)^*\otimes H)\otimes J @>>\ass((G\otimes F)^*,H,J)>(G\otimes 
F)^*\otimes(H\otimes J)
\endCD$$

\definition{Definition \propno} The set of standard iterated functors is the
smallest set of iterated functors which satisfies:
\roster
\item The iterates $e$, $1$ and $*$ are standard.
\item If $F\ne e$ is standard then $F\otimes 1$ and $F\otimes *$ are standard.
\endroster  
\enddefinition 

\definition{Definition \propno} The set of reduced iterated functors is the
smallest set of iterated functors which satisfies:
\roster
\item The iterates $e$, $1$ and $*$ are reduced.
\item If $F$ and $G$ are reduced then $F\otimes G$ is reduced.
\endroster  
\enddefinition         

Define two iterated functors, $F$ and $G$, to be equivalent if there is an
iterated natural transformation from $F$ to $G$.
This is obviously an equivalence relation.

\definition{Definition \propno} The rank $r$ and luminosity, $l$, of an 
iterated functor are defined recursively as follows: 
\roster 
\item $r(e)=1$ and $r(1)=1$
\item $r(F\otimes G)=r(F)+r(G)+1$
\item $r(F^*)=r(F)$
\endroster
\roster 
\item $l(e)=0$ and $l(1)=0$. 
\item $l(F\otimes G)=l(F)+l(G)$. 
\item $l(F^*)=l(F)+r(F)$. 
\endroster 
\enddefinition 
The rank is a modified multiplicity, and the luminosity is a measure of the 
number of stars. The luminosity is unaffected by central iterated natural 
transformations, but is increased by instances of $\pdual$, $\nat$ and $\pv$. 
                                                             
\proclaim{Lemma \propno\name\reduced} For any iterated functor $F$, there is a 
unique reduced iterated functor $R$ and a unique natural transformation $R\to 
F$ which is an iterate of instances of $\pdual$, $\nat$ and $\pv$. 
\endproclaim 

\demo{Proof} The proof is by induction on the sum of the luminosity and the 
rank of $F$. Any iterated functor is of the form
\roster
\item $F^{**}$.
\item $(F\otimes G)^*$.
\item $e^*$
\item $F\otimes G$.   
\item $1$, $*$, or $e$.
\endroster
and the only instances of $\pdual$, $\nat$ and $\pv$ which have these as range 
are
\roster
\item $\nat\colon F\to F^{**}$
\item $\pdual\colon (G^*)\otimes(F^*)\to(F\otimes G)^*$
\item $\pv\colon e\to e^*$
\item $A\otimes 1$ or $1 \otimes B$
\item none.
\endroster
In the first three cases, the conclusion follows from the induction hypothesis 
applied to the domain of the natural transformation, which has the same rank 
and lower luminosity. In (4), the induction hypothesis can be applied to $F$ 
and $G$ separately, which have lower rank and luminosity which is no greater.
The conclusion follows as the natural transformations $A\otimes 1$ and $1 
\otimes B$ commute. The functors in (5) are reduced iterated functors, and 
$F=R$.
\enddemo

The proof of the coherence theorem \coherence\ now follows. It is only 
necessary to consider reduced iterated natural transformations.  
A reduced iterated natural transformation is said to be directed if it is the 
product of instances which do not include the inverses of $(\pdual, \nat, 
\pv)$.  By the preceding lemma \reduced, each iterated functor is equivalent 
to a reduced iterated functor, and hence to a standard iterated functor, by a 
directed natural transformation.

By a standard argument (see \cite{MacLane  1963}),
it is sufficient to prove the coherence result for directed natural 
transformations between an iterated functor $F$ and a standard iterated 
functor $S$.
According to lemma \directed, any directed natural transformation $S\to F$ can 
be written as a central iterate $S\to R$ followed by an iterate of instances 
of $(\pdual, \nat, \pv)\colon R\to F$. By lemma \reduced, $R$ and the second 
natural transformation are unique, whilst the first natural transformation is 
unique by the coherence of central iterates.

\definition{Definition \propno} If $\Cal C$ is a category with duals
then define the category $P{\Cal C}$ to have objects of the form
$(F,\{a_1,\ldots ,a_n\})$ where $F$ is a standard iterated functor
 of multiplicity
$n$ and each $a_i$ is an object of $\Cal{C}$. The set of morphisms from
$(F,\{a_1,\ldots ,a_n\})$ to $(G,\{a_1,\ldots ,a_m\})$ is the set of
morphisms in $\Cal C$ from $F(a_1,\ldots ,a_n)$ to $G(a_1,\ldots ,a_m)$. 
\enddefinition

There is a functor $i\colon {\Cal C}\to P{\Cal C}$ defined on 
objects by $a\mapsto (1,a)$. There is also a functor 
$\pi\colon P{\Cal C}\to {\Cal C}$ defined on objects by
$(F,\{a_1,\ldots ,a_n\}) \mapsto F(a_1,\ldots ,a_n)$. It is trivial
that the composite $i\circ\pi$ is the identity functor of $\Cal C$.
It is also clear that there is a canonical natural transformation from
$\pi\circ i$ to the identity functor of $P{\Cal C}$. Hence the categories
$P{\Cal C}$ and ${\Cal C}$ are canonically equivalent.

The coherence theorem implies that the category $P{\Cal C}$ is a strict
monoidal category with strict duals. The functor $i$ has the universal
property that if $\Cal D$ is a strict monoidal category with strict duals and 
$F\colon{\Cal C}\to{\Cal D}$ is a functor of monoidal categories with duals
then there is a unique functor $G\colon P{\Cal C}\to{\Cal D}$ of strict
monoidal categories with strict duals such that $F=i\circ G$. 

\head\headno Spherical categories \endhead

In this section we introduce spherical categories. A pivotal category
is a category with duals and with some additional structure and a
spherical category is a pivotal category in which the additional structure
satisfies an extra condition.   

Strict pivotal categories are discussed  in \cite{Freyd and Yetter  1989}, where it is shown
that the category of oriented planar graphs up to isotopy with labelled edges
and with a distinguished edge at each vertex is a strict pivotal category. 

\nextfigure\trace
If the sphere $S^2$ is regarded as the plane with the point at infinity
attached, then a closed graph in the plane can be regarded as a closed graph
on the sphere. There is an isotopy of the sphere which takes a closed graph
of the form of Figure \trace\ to the graph obtained by closing $M$ in a loop
to the left. This isotopy moves the loop in Figure \trace\  past the point at
infinity. Taken together with planar isotopies, such an operation on planar
graphs generates all the isotopies on the sphere. It follows that the
evaluation in a spherical category of a closed graph with labelled edges and
with a distinguished edge at each vertex is invariant under isotopies of the
sphere $S^2$.

\definition{Definition \propno} 
A pivotal category is a category with duals together with a morphism
$\epsilon (c)\colon e\to c\otimes c^*$ for each object $c\in{\Cal C}$.

The conditions on the components of $\epsilon$ are the following:
\roster
\item For all morphisms, $f\colon a\to b$, the following diagram commutes
$$\CD
e @>{\epsilon (a)}>> a\otimes a^*\\
@V{\epsilon (b)}VV @VV{f\otimes 1}V\\
b\otimes b^* @>>{1\otimes f^*}> b\otimes a^*
\endCD$$
\item For all objects $a$, the following composite is the identity map
of $a^*$:
$$\multline a^* @>{\lambda^{-1}(a^*)}>> e\otimes a^* @>{\epsilon (a^*)\otimes 1}>>
(a^*\otimes a^{**})\otimes a^* @>{\ass (a^*,a^{**},a^*)}>> a^*\otimes (a^{**}
\otimes a^*) @>{1\otimes\gamma(a^*,a)}>> \\ 
a^*\otimes (a\otimes a^*)^* @>{1\otimes \epsilon(a)^*}>> 
a^*\otimes e^* @>1\otimes\pv^{-1}>>a^*\otimes e @>\rho(a^*)>>
a^*\endmultline$$
\item For all objects $a$ and $b$ the following composite is required to
be $\epsilon (a\otimes b)$:
$$\multline e @>{\epsilon (a)}>> a\otimes a^* @>{1\otimes\lambda^{-1} (a^*)}
>> a\otimes (e\otimes a^*) \\
@>{1\otimes (\epsilon (b)\otimes 1)}>>
a\otimes ((b\otimes b^*)\otimes a^*) @>{1\otimes \ass (b,b^*,a^*)}>>
a\otimes (b\otimes (b^*\otimes a^*)) \\ 
@>{\ass^{-1}(a,b,b^*\otimes a^*)}>>
(a\otimes b)\otimes (b^*\otimes a^*) @>1\otimes{\pdual (b,a)}>>
(a\otimes b)\otimes (a\otimes b)^*\endmultline$$
\endroster
\enddefinition

These conditions imply that the strict extension is a
pivotal category in the sense of \cite{Freyd and Yetter  1989}. This is clear once it
has been shown that $\epsilon$ is well-defined on the strict extension.

The functor $*$ and the maps $\epsilon$ are not independent.  The maps
$\epsilon$ determine $*$. This observation is in \cite{Kelly and Laplaza  1980}  where the
maps $\epsilon$ are taken as more fundamental than $*$.
In \cite{Turaev  1992} 
it is noted that the maps $\epsilon$ also determine $\pdual$.

\proclaim{Lemma \propno\name\dual}
In any pivotal category, for any morphism
$f\colon a\to b$ the following composite is $f^*$:
$$\multline
{b^*} @>{\rho^{-1} ({b^*})}>> {b^*}\otimes e
@>{1\otimes\epsilon (a)}>> {b^*}\otimes (a\otimes{a^*})
@>{1\otimes (f\otimes 1)}>> {b^*}\otimes (b\otimes{a^*})\\
@>{\ass^{-1}({b^*},b,{a^*})}>> ({b^*}\otimes b)\otimes {a^*}
@>{(\nat(b^*)\otimes\nat(b))\otimes 1}>>(b^{***}\otimes b^{**})\otimes a
@>{\pdual(b^{**},b^*)\otimes 1}>>(b^*\otimes b^{**})^*\otimes a^*\\
@>{\epsilon^* (b^*)\otimes 1}>> e^*\otimes{a^*}
@>{\pv^{-1}\otimes 1}>>e\otimes a^* 
@>{\lambda({a^*})}>> {a^*}\endmultline$$
\endproclaim

\demo{Proof}  This follows directly from conditions (1) and (2) of
the preceding definition.
\enddemo

\remark{Remark \propno\name\adjoint}
Note also that a pivotal category is closed, in the sense that there are
natural isomorphisms, for all $a$, $b$ and $c$;
$$\matrix
\Hom (a\otimes b,c)\cong\Hom (b,a^*\otimes c) &
\Hom (a,b\otimes c)\cong\Hom (a\otimes c^* ,b) \\
\Hom (a\otimes b,c)\cong\Hom (a,c\otimes b^*) &
\Hom (a,b\otimes c)\cong\Hom (b^*\otimes a,c) 
\endmatrix$$\endremark

\goodbreak\midinsert
\centerline{\epsfbox{trace.eps}}
\figno\endinsert

\definition{Definition \propno} Let $a$ be any object in a 
pivotal category. Then the monoid $\End (a)$ has two trace maps, 
$\tr_L,\tr_R\colon\End (a)\to\End (e)$.
In a pivotal category $\tr_L(f)$ is defined to be the composite
$$\multline
e @>{\epsilon (a^*)}>>
a^*\otimes a^{**} @>{1\otimes\nat^{-1}(a)}>>
a^*\otimes a @>{1\otimes f}>>
a^*\otimes a @>{\nat(a^*)\otimes\nat (a)}>> \\
a^{***}\otimes a^{**} @>{\pdual (a^{**},a^*)}>>
(a^*\otimes a^{**})^* @>{\epsilon^*(a^*)}>> e^* @>{\pv}>> e
\endmultline$$
and $\tr_R(f)$ is defined to be the composite
$$\multline
e @>{\epsilon (a)}>>
a\otimes a^* @>{f\otimes 1}>>
a\otimes a^* @>{\nat (a)\otimes 1}>> 
a^{**}\otimes a^* @>{\pdual (a^*,a)}>>
(a\otimes a^*)^* @>{\epsilon^*(a)}>>
e^* @>{\pv}>> e
\endmultline$$
In a strict pivotal category these definitions simplify to:
$$\align \tr_L(f)&=\epsilon (a^*)(1\otimes f)\epsilon (a^*)^*\\
\tr_R(f)&=\epsilon (a)(f\otimes 1)\epsilon (a)^*\endalign$$
These are called trace maps because they satisfy $\tr_L(fg)=\tr_L(gf)$
and $\tr_R(fg)=\tr_R(gf)$.
\enddefinition

\definition{Definition \propno} A pivotal category is spherical if,
for all objects $a$ and all morphisms $f\colon a\to a$,
$$\tr_L(f)=\tr_R(f).$$
\enddefinition

An equivalent condition is that $tr_L(f)=tr_L(f^*)$, for all 
$f\colon a\to a$. 
Also, in a spherical category, $\tr_L(f\otimes g)=\tr_L(f).\tr_L(g)$
(where the product is in $\End(e)$)
for all $f\colon a\to a$ and all $g\colon b\to b$.

In the rest of this paper we study additive spherical categories.
This means that each $\Hom$ set is a finitely generated abelian group;
and that the data defining the spherical structure is compatible with the 
additive structure. This means that $\otimes$ is $\Z$-bilinear, that $*$
is $\Z$-linear and all components of all the natural transformations in the
pivotal data are multilinear over $\Z$. 

In any additive monoidal category $\End (e)$ is a commutative ring
(see \cite{Kelly and Laplaza  1980})  and we denote this ring by $\ring$. 
In particular the trace map, $\tr_L$, takes values in this ring.
It follows that an additive monoidal category is $\ring$-linear, by using
the structure maps of $\lambda$ and $\rho$, and that
if the category is pivotal then the data defining the pivotal structure
is $\ring$-linear.

The main examples of additive spherical categories arise
as the category of representations of a Hopf algebra with some additional
structure. This is discussed in detail in the next section.

\example{Example \propno} An example of an additive spherical category 
which cannot be regarded as a category whose objects are finitely
generated free modules is given by taking the free $\Z [q,q^{-1},z]$-linear
category on the category of oriented tangles and then taking the quotient
by the well-known skein relation for the HOMFLY polynomial. This is an
additive spherical category and for each pair of objects $a$ and $b$,
$\Hom (a,b)$ is a finitely generated free $\Z [q,q^{-1},z]$-module.
However the objects cannot be taken to be finitely generated modules unless
$z$ is a quantum integer.
\endexample

\definition{Definition \propno} For any two objects $a$ and $b$ there
is a bilinear pairing
$$\Theta\colon\Hom (a,b)\times\Hom (b,a)\to\ring$$
defined by $\Theta (f,g)=\tr_L(fg)=\tr_L(gf)$.
\enddefinition

\definition{Definition \propno}
A spherical category is non-degenerate if, for all objects $a$ and $b$,
the pairing $\Theta$ is non-degenerate. \enddefinition

The next theorem shows that every additive spherical category has a
natural quotient which is a non-degenerate spherical category.
This construction is mentioned in \cite{Turaev  1992}.
\proclaim{Theorem \propno \name\quotient} Let ${\Cal C}$ be an additive 
spherical category. Define the subcategory ${\Cal J}$ to have the same set of 
objects and morphisms defined by
$$\Hom_{\Cal J}(c_1,c_2)=\{f\in \Hom_{\Cal C}(c_1,c_2): \tr_L (fg)=0\qquad
\text{for all $g\in \Hom_{\Cal C}(c_2,c_1)$}\}$$
Then ${\Cal C}/{\Cal J}$ is a non-degenerate additive spherical category.
\endproclaim
\demo{Proof} It is clear that ${\Cal J}$ is closed under composition on either
side by arbitrary morphisms in ${\Cal C}$. Hence the quotient is an
additive category. It is also clear that $f\in {\Cal J}$ if and only if $f^*\in {\Cal J}$
and so the functor $*$ is well-defined on the quotient. The functor
$\otimes$ is well-defined on the quotient since $f\in {\Cal J}$ implies 
$f\otimes g_1\in {\Cal J}$ and $g_2\otimes f\in {\Cal J}$ for arbitrary morphisms in
${\Cal C}$. This follows from the observation that $\tr_L((f\otimes g)h)$ can always be written as $\tr_L(f h')$ for a morphism $h'$, which can be constructed from $g$ and $h$.\footnote{We thank Marco Mackaay for pointing out that this argument was given incorrectly in \cite{Barrett and Westbury 1996}.}

The structure maps of the natural transformations $\ass$, $\lambda$, $\rho$,
$\pdual$, $\pv$, $\tau$ and the morphisms $\epsilon (a)$ are taken to be
the images in the quotient of the given morphisms in ${\Cal C}$. The
conditions on this structure which imply that this quotient is spherical
follow from the same conditions in ${\Cal C}$.

Each pairing $\Theta$ is non-degenerate by construction.
\enddemo
\head\headno  Spherical Hopf algebras \endhead
In this section we introduce spherical Hopf algebras. These are defined
to be Hopf algebras with some additional structure which implies that the
category of finitely generated modules is a spherical category.

\definition{Definition \propno \name\sphericalHopf} A spherical Hopf algebra
over $\ring$ consists of a $\ring$-module $A$
together with the following data \roster
\item a multiplication $\mu$
\item a unit $\eta\colon \ring\to A$
\item a comultiplication $\Delta\colon A\to A\otimes A$
\item a counit $\epsilon\colon A\to \ring$
\item an antipodal map $\gamma\colon A\to A$
\item an element $w\in A$
\endroster\enddefinition

The data $(A,\mu ,\eta ,\Delta ,\epsilon ,\gamma )$ is required to define a
Hopf algebra. The conditions on the element $w$ are the following:
\roster
\item $\gamma^2(a)=waw^{-1}$ for all $a\in A$.
\item $\Delta (w)=w\otimes w$.
\item $\gamma (w)=w^{-1}$.
\item $\epsilon(w)=1$.
\item $\tr (\theta w)=\tr (\theta w^{-1})$ for any left $A$-module, $V$, which is
finitely generated and projective as an $\ring$-module and for all $\theta\in\End_A(V)$.
\endroster
It follows from the condition $\Delta(w)=w\otimes w$ that $\gamma (w)=w^{-1}$
and that $\epsilon(w)=1$. Such elements are called group-like.

\example{Example \propno} Examples of Hopf algebras which are spherical are:
\roster
\item Any involutory Hopf algebra is spherical. The element $w$ can be taken
to be 1.
\item Any ribbon Hopf algebra, as defined in \cite{Reshetikhin and Turaev  1990},
is spherical. The element $w$ can be taken to be $uv^{-1}$ where the
element $u$ is determined by the quasi-triangular structure and the
element $v$ is the ribbon element.
\endroster\endexample

\remark{Remark \propno} A Hopf algebra with an element $w$ that satisfies
the first two conditions of definition \sphericalHopf\  is spherical if, 
either $w^2=1$, or all modules are self-dual.\endremark

\remark{Remark \propno} If $A$ is a Hopf algebra there may exist more
than one element $w$ such that $(A,w)$ is a spherical Hopf algebra.
However, if $w_1$ is one such element then $w_2=gw_1$ is another such
element if and only if $g$ satisfies the conditions:
\roster
\item $g$ is central
\item $g$ is group-like
\item $g$ is an involution
\endroster
In particular, if $A$ is finite dimensional over a field, the set of $g$ 
satisfying these conditions form a group of the form $\Z_2^k$ for some 
integer $k$.
\endremark

\example{Example \propno} This is an example of a finite dimensional
Hopf algebra over $\C$ which satisfies all the conditions for a spherical 
Hopf algebra except that the left and right traces are distinct. This 
example is the quantised enveloping algebra of the Borel subalgebra of 
$SL_2(\C)$.

Let $s$ be a primitive $2r$-th root of unity
with $r>1$. Let $B$ be the unital algebra generated by
elements $X$ and $K$ subject to the defining relations
$$\align
KX&=sXK\\
K^{4r}&=1\\
X^r&=0
\endalign$$

Then $B$ is a finite dimensional algebra and also has a Hopf algebra
structure defined by:
\roster
\item The coproduct, $\Delta$, is defined by
$\Delta (K)= K\otimes K$ and $\Delta (X)= X\otimes K + K^{-1}\otimes X$
\item The augmentation, $\epsilon$, is defined by
$\epsilon (K)=1$ and $\epsilon (X)=0$
\item The antipode, $\gamma$, is defined by
$\gamma (K)= K^{-1}$ and $\gamma (X)= -sX$
\endroster

The element $w=K^2$ satisfies the conditions
$$\align
\Delta (w)&=w\otimes w\\
\epsilon (w)&=1\\
\gamma (w)&=w^{-1}\\
\gamma^2 (b)&=wbw^{-1}\quad\text{for all $b\in B$}
\endalign$$

The trace condition is not satisfied since $B$
has $4r$ one dimensional representations with $X=0$ and $K$ a $4r$-th
root of unity and it is clear that the trace condition is not satisfied
in these representations.
\endexample

In the rest of this paper a left $A$-module will mean a left $A$-module which is
finitely generated and free as a $\ring$-module.
\proclaim{Proposition \propno} If $A$ is a spherical Hopf algebra over $\ring$ then
the category of left $A$-modules is a spherical category.
\endproclaim
\demo{Proof}
The category of modules of a bialgebra is monoidal. The category of
modules of a Hopf algebra has a functor $*$ which sends a
module to its contragradient dual and a morphism to its adjoint.

An element $w$ in a Hopf algebra which satisfies conditions (1) and (2)
determines a pivotal structure for the category of modules. For each
module $a$, the map $\nat (a)$ given by the natural transformation
$\nat\colon 1 \to **$ is defined to be the natural identification of $\ring$-modules 
$a\to a^{**}$ followed by multiplication by $w^{-1}$. This is a morphism of
modules. The morphism $\epsilon_a$ sends 1 to the canonical element of
$a\otimes a^*$.

The two traces are given by $\tr_L(\theta)=\tr (\theta w)$ and $\tr_R(\theta)=
\tr(\theta w^{-1})$. This shows that the category of modules of a
spherical Hopf algebra is an additive spherical category.\enddemo

The rest of the paper is a discussion of the properties of a spherical
category which is constructed by taking the non-degenerate quotient of
a spherical subcategory of the category of modules of a spherical
Hopf algebra. In the rest of this paper we assume that $\ring$ is an 
algebraically closed field. These results show that this quotient
is semisimple. If the Hopf algebra is itself semisimple this is clear.

\proclaim{Proposition \propno} Let $A$ be a spherical Hopf algebra.
Let $x$ be a left $A$-module. Then $\End (x)$, the
endomorphism algebra in the non-degenerate quotient, is semi\-simple.
\endproclaim
\demo{Proof} Let $\End_A(x)$ be the endomorphism algebra of $x$ in the
category of left $A$-modules. Then we show that
the kernel of the map $\End_A(x)\to\End(x)$ contains the
nilpotent radical of $\End_A(x)$. Since the nilpotent radical is a two-sided
ideal it is sufficient to show that if $\theta\in\End_A(x)$ is nilpotent
then $\tr_L(\theta)=0$. However $\theta$ and $\omega$ commute and so
$\theta\omega$ is also nilpotent. This shows that $\tr_L(\theta)$ is the
matrix trace of a nilpotent endomorphism of a finite dimensional vector
space over $\ring$ and so $\tr_L(\theta)=0$.
\enddemo

\proclaim{Proposition \propno} Let $A$ be a spherical Hopf algebra.
Take the non-degenerate quotient of the spherical category of
left $A$-modules. Take $J$ to be a set of objects such
that, if $\End (x)\cong\ring$ then there is a unique $a\in J$ such that
$x\cong a$. Then, for each object $x$ in the non-degenerate quotient,
there is a natural isomorphism
$$x\cong\bigoplus_{a\in J}\Hom (x,a)\otimes a .$$
\endproclaim

\demo{Proof}
First we define, for each idempotent $\pi\in\End (x)$, an object $\pi x$
in the non-degenerate quotient. Let $\pi^\prime\in\End_A(x)$ be a lifting
of $\pi$. Then $\pi^\prime x$ is a left $A$-module and so also gives an
object, $\pi x$, in the non-degenerate quotient. This object is independent
of the choice of lifting since $\pi x$ is characterised by the property
that, for any $w$ and $y$,
$$\Hom (w,\pi x)=\Hom (w,x).\pi \quad\text{ and }\quad
\Hom (\pi x,y)=\pi .\Hom (x,y) .$$

Next we show that if $\pi$ is primitive then $\End (\pi x)\cong\ring$.
More generally, if $\pi_1$ and $\pi_2$ are primitive idempotents then
$$\Hom (\pi_1x,\pi_2x)\cong\cases
\ring&\text{if $\pi_1$ and $\pi_2$ are isomorphic}\\
0&\text{otherwise}\endcases$$
If $\pi$ is primitive then so is $\pi^\prime$ and so $\End_A(\pi^\prime x)$
is a local algebra. Hence $\End (\pi x)$ is a non-trivial semisimple
quotient of a local algebra and so is $\ring$. Alternatively, $\End (\pi x)$
is $\pi.\End (x).\pi$ which is $\ring$ since $\pi$ is primitive.

Now write $1\in\End (x)$ as a sum of orthogonal primitive idempotents.
This decomposition is not canonical. However, for each $a\in J$, define
$E_a$ to be the sum of the idempotents, $\{\pi\}$, in the decomposition
that satisfy $\pi x\cong a$. Then each non-zero $E_a$ is a central
idempotent in $\End (x)$ and the decomposition
$$x\cong\bigoplus_{a\in J}E_ax$$
is canonical. Furthermore it follows from this discussion that, for each
$a\in J$, $E_ax$ can be naturally identified with $V(x,a)\otimes a$ for
some $V(x,a)$. Hence $x\cong\bigoplus_{a\in J}V(x,a)\otimes a$. 

Each $V(x,a)$ can be identified with $\Hom (x,a)$ since $\Hom (a,b)=0$ 
if $a$ and $b$ are distinct elements of $J$. This follows from the 
non-degeneracy condition which implies that if $\Hom(a,b)=0$ then 
$\Hom(b,a)=0$.
\enddemo

This shows that the non-degenerate quotient is semisimple and that the set
$J$ is a complete set of inequivalent simple objects. If the Hopf algebra
is finite dimensional and semisimple then it is clear that this set is
finite. If the Hopf algebra is finite dimensional but not semisimple then
the set $J$ is typically infinite.

There are two 3-manifold invariants constructed from the quantised
enveloping algebra of $SL(2)$. One is the construction of a invariant
of a framed 3-manifold given a surgery presentation of the 3-manifold
given in \cite{Reshetikhin and Turaev  1991}. The other is a state sum model invariant
constructed in \cite{Turaev and Viro  1992}  using a triangulation. The formalism for
extending the definitions from $SL(2)$ to other finite dimensional simple
algebras is given in \cite{Turaev  1992}  and \cite{Barrett and Westbury 1996}.
Each of these approaches
encounters the same difficulty which is that the set $J$ is required to be
finite. A quantised enveloping algebra at a root of unity is not
semisimple and the set $J$ for the category of all  
modules may be infinite. In order to define 3-manifold invariants it
is sufficient to find a spherical subcategory with $J$ finite.
This problem was solved for $SL(2)$ in \cite{Reshetikhin and Turaev  1991}. 

Let $A$ be the quantised enveloping algebra of a finite dimensional
semisimple Lie algebra at a root of unity. Assume that the order, $k$, of
the quantum parameter $q$ is at least the Coxeter number of the Lie
algebra. Then it is shown in \cite{Andersen and Paradowski 1995}  
that the category of tilting modules is the smallest category which
contains the irreducibles with highest weight in the interior of the
fundamental alcove and which is closed under taking tensor products and
direct summands. Furthermore this category is spherical and the isomorphism
classes of objects in the semisimple quotient are indexed by the dominant
weights in the interior of the fundamental alcove, and in particular, is
a finite set.

\remark{Remark \propno\name\hopf} Let $A$ be a finite dimensional
spherical Hopf algebra. Then the category of projective
left $A$-modules is a spherical category. This constructs a spherical
subcategory with $J$ finite but we know of no applications. If $A$ is
semisimple this result is superfluous and in the example of the quantised
enveloping algebra of $SL(2)$ given in \cite{Reshetikhin and Turaev  1991}  every indecomposable
projective has quantum dimension 0 so that $J$ is empty.

The category of projective left $A$-modules
is a subcategory of the spherical category of left
$A$-modules. Hence to show it is spherical it is sufficient to show it is
closed under the operations of tensor product and taking duals.
The tensor product of a projective module with any module is projective
and so the category of projective modules is closed under tensor products.
It is shown in \cite{Larson and Sweedler  1969}  that a finite dimensional Hopf algebra is
Frobenius and therefore the dual of a projective module is projective.
\endremark

\enddocument